\documentclass[aps,prmaterials,reprint,amsmath,amssymb,superscriptaddress,longbibliography]{revtex4-2}

\usepackage{graphicx}
\usepackage{booktabs}
\usepackage{xcolor}
\usepackage[colorlinks=true,linkcolor=blue,citecolor=blue,urlcolor=blue]{hyperref}
\usepackage{multirow}
\usepackage{subcaption}

\newcommand{\safefig}[2]{%
  \IfFileExists{#1}{\includegraphics[width=#2]{#1}}%
  {\fbox{\parbox{0.9\linewidth}{\centering\small\texttt{Missing figure file: #1}}}}}

\newcommand{\Nten}{N_{10\%}}
\newcommand{\RMSEfull}{\mathrm{RMSE}_{\mathrm{full}}}

\begin{document}

\title{Systematic Evaluation of Training Data Generation Strategies for Machine Learning-Based Molecular and Materials Property Prediction}

\author{Minh Nghia Nguyen}
\affiliation{Institute of Chemistry and Biochemistry, Freie Universit\"at Berlin, Arnimallee 22, 14195 Berlin, Germany}


\author{Beate Paulus}
\affiliation{Institute of Chemistry and Biochemistry, Freie Universit\"at Berlin, Arnimallee 22, 14195 Berlin, Germany}

\author{Akhil S. Nair}
\thanks{Corresponding author: akhil.sugathan.nair@fu-berlin.de}
\affiliation{Department of Polymer and Functional Materials, CSIR-Indian Institute of Chemical Technology, 500007, Hyderabad, India}

\date{\today}

\begin{abstract}
Machine-learning (ML) models for molecular and materials property prediction are often trained on large datasets curated through arbitrary or generic sampling strategies. Such approaches can lead to the over- or under-representation of regions of chemical and materials space that are most relevant to the prediction task. An important challenge is therefore how to efficiently curate training data to maximize predictive performance while minimising the number of labelled examples required. Here, we systematically compare training data acquisition strategies for molecular and materials property prediction, ranging from random sampling to various active learning approaches. We evaluate their performance across different model classes, representations, datasets, and target properties. Our results show that the effectiveness of data acquisition strategies depends strongly on the interplay between the model, dataset, and prediction task, although uncertainty-based sampling generally outperforms alternative strategies. Based on these findings, we derive guidelines for selecting data acquisition strategies under a given labelling budget.

\end{abstract}

\maketitle

\section{Introduction}
\label{sec:intro}

Machine learning (ML) can accelerate molecular and materials discovery by predicting properties that would otherwise require expensive experiments or high-fidelity simulations. The predictive performance of ML models depends not only on their architectures but also on the quality, diversity, and composition of the training data. However, datasets are often constructed through exhaustive or near-random sampling of chemical space, leading to substantial redundancy among chemically or structurally similar samples. Such redundancy increases both data-generation and model-training costs without necessarily yielding commensurate improvements in predictive accuracy. Recent studies have demonstrated that large molecular and materials datasets can be substantially reduced while retaining much of their predictive performance, although the extent of redundancy depends on the target property, dataset, and representation~\cite{Li2023,Li2024,moerman2026atoms}.

Active learning (AL) provides a systematic alternative to passive dataset construction by iteratively selecting samples for labeling according to criteria such as predictive uncertainty, structural diversity, or their combination. This approach is particularly valuable in atomistic modelling, where obtaining electronic-structure labels is computationally expensive and adequately sampling diverse chemical environments remains challenging~\cite{Kulichenko2024}. Nevertheless, the effectiveness of sampling strategies used in AL to acquire the data is not universal; it can vary with the dataset, target property, model class, data representation, and acquisition criterion~\cite{Bi2025, Nair2026}. Consequently, identifying an efficient data-generation strategy for a given prediction task requires understanding how these factors interact. Despite growing interest in data-efficient learning, a systematic assessment of their individual and combined effects across different model classes, representations, and molecular and materials datasets remains limited.

In this work, we investigate three central questions: (i) To what extent are widely used molecular and materials datasets redundant across different ML model classes? (ii) Which sampling strategies reduce labelling requirements most effectively while preserving predictive accuracy? and (iii) How does the data generation efficiency depend on the representation used to describe molecules and materials? To address these questions, we benchmark conventional ML and deep-learning (DL) models across datasets spanning molecular systems and inorganic crystals, considering dataset sizes ranging from approximately \(10^3\) to \(10^5\) samples. By systematically comparing alternative sampling strategies, model classes, and data representations, we quantify the trade-offs between training-set size, labeling effort, and predictive performance. This analysis provides a comparative assessment of data efficiency across diverse learning settings and identifies factors that govern the effectiveness of training-data generation methods for molecular and materials property prediction.

\section{Methods}
\label{sec:methods}

\subsection{Datasets}

\begin{figure*}[t]
\centering
\begin{subfigure}{0.48\linewidth} 
    \centering
    \safefig{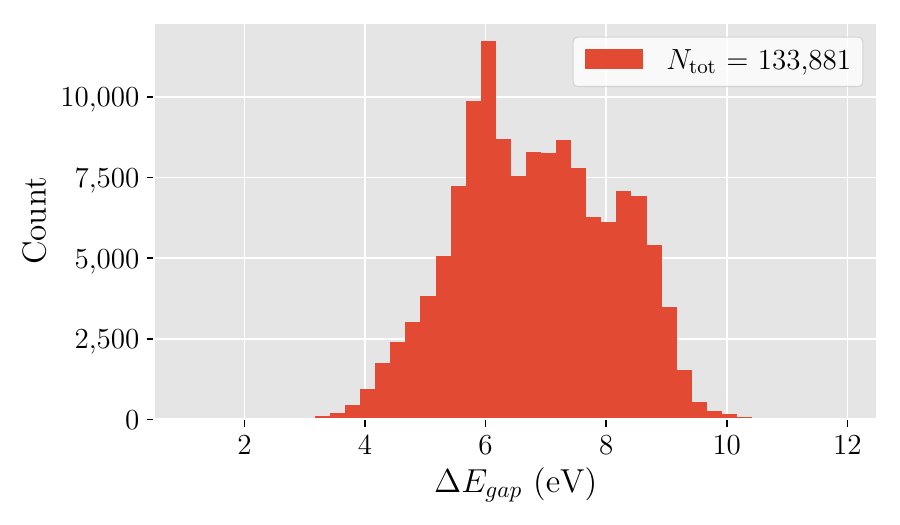}{\linewidth}
    \caption{}
\end{subfigure}
\hfill 
\begin{subfigure}{0.48\linewidth} 
    \centering
    \safefig{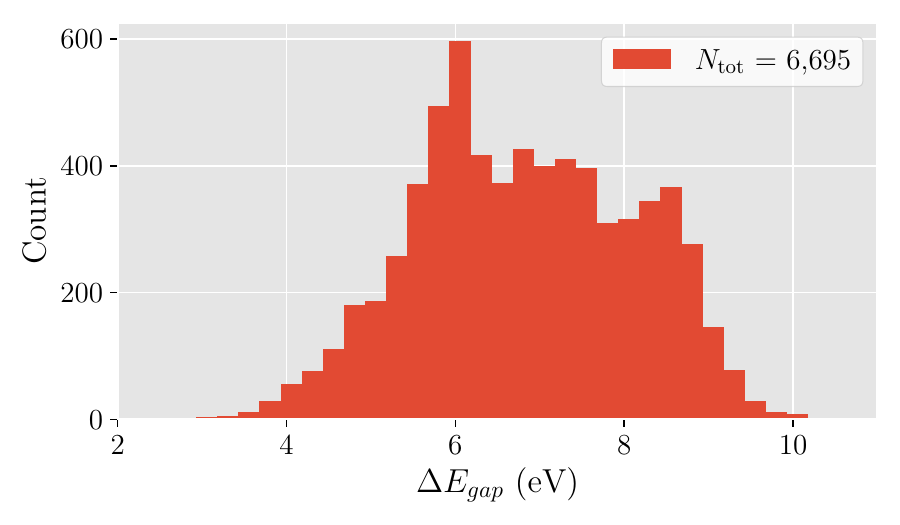}{\linewidth}
    \caption{}
\end{subfigure}
\vfill
\begin{subfigure}{0.48\linewidth} 
    \centering
    \safefig{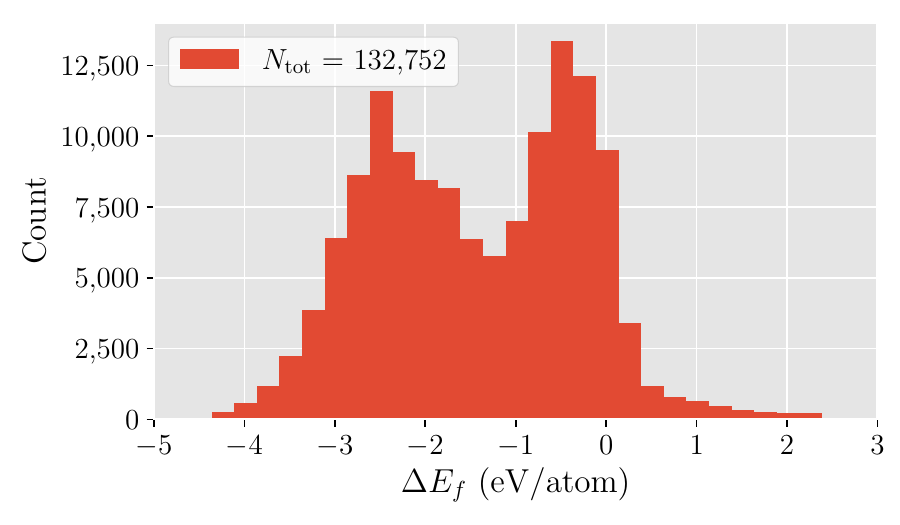}{\linewidth}
    \caption{}
\end{subfigure}
\hfill 
\begin{subfigure}{0.48\linewidth} 
    \centering
    \safefig{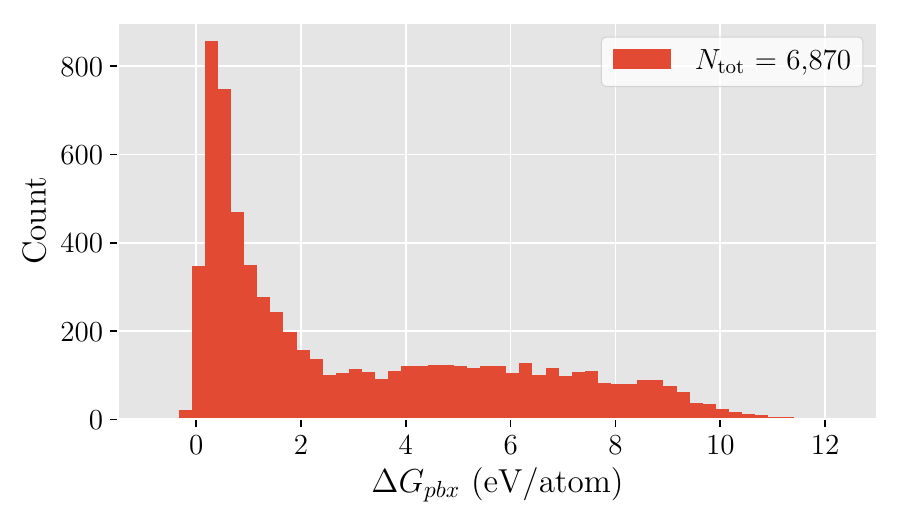}{\linewidth}
    \caption{}
\end{subfigure}
\caption{Target property distributions of \textbf{(a)} QM9, \textbf{(b)} QM9small, \textbf{(c)} the Matbench MP and \textbf{(d)} the Pourbaix decomposition free energy dataset. The legend in each panel reports the total number of samples $N_{\mathrm{tot}}$. All histograms use a constant bin width of $0.25$ in the units of the respective x-axis. Symbols on the x-axes denote the target properties: $\Delta E_{gap}$, the HOMO-LUMO gap (eV); $\Delta E_f$, the formation energy per atom (eV/atom); and $\Delta G_{pbx}$, the Pourbaix decomposition free energy per atom (eV/atom).}
\label{fig:distributions}
\end{figure*}

For molecular property prediction, we consider the QM9 dataset~\cite{Ramakrishnan2014}, comprising 133,881 small organic molecules after removing duplicate structures and invalid entries. The target property is the HOMO–LUMO gap ($\Delta E_{gap}$), calculated at the B3LYP level of theory. To probe the same domain at a much smaller scale, we use \texttt{QM9small}, a random 5\% subset of QM9 containing 6,695 molecules with the same target property. For materials property prediction, we use the Matbench Materials Project (MP) dataset, which contains 132,752 inorganic crystals, with the DFT-PBE formation energy per atom ($\Delta E_f$) as the target property~\cite{Jain2013, Dunn2020}. As a smaller dataset for materials property prediction, we consider the Pourbaix Stability dataset introduced in our previous study~\cite{Nair2025}, comprising 6,870 inorganic crystals with DFT-HSE06-calculated Pourbaix decomposition free energies ($\Delta G_{pbx}$ in eV/atom) as the target property. The four datasets thus form two large--small pairs, one for molecules and one for materials, which allows the effect of dataset size to be examined alongside that of the target property. The distributions of the target properties across all four datasets are shown in Figure~\ref{fig:distributions}.

\subsection{Machine Learning Models and Representations}
For classical ML models, we employ descriptor-based representations to encode molecules and materials. For the QM9 dataset, molecules are represented by eight scalar properties such as $\mu$ (dipole moment), $\alpha$ (isotropic polarizability), ZPVE (zero-Point vibrational Energy), $U_0$, $U_{298}$ (internal energies at 0 and 298.15~K)  $H_{0}$, $H_{298}$ (enthalpies at 0 and 298.15 K), $G_{298}$ (Gibbs free energy at 298.15 K) and $C_v$ (heat capacity at 298.15~K). These are further combined with Morgan fingerprints (radius of 2 and 2\,048 bits)~\cite{Morgan1965, Rogers2010} and MACCS keys~\cite{Durant2002}, generated using RDKit~\cite{RDKit}. For the Matbench MP dataset, we use composition-weighted elemental descriptors from the Magpie feature set~\cite{Ward2016}, implemented through matminer~\cite{Ward2018}. For the Pourbaix stability dataset, we use the feature set provided as in Ref.~\cite{Nair2025}, which augments the Magpie features with additional elemental descriptors.

We consider three classical ML algorithms: random forests (RF)~\cite{Breiman2001}, kernel ridge regression (KRR) with a Laplacian kernel, and extreme gradient boosting (XGB)~\cite{Friedman2001}. For KRR, we employ the Nyström approximation~\cite{Williams2000} for the two larger datasets to reduce computational costs. The QM9small and Pourbaix stability datasets use exact KRR. The predictions from the three models are combined using a voting regressor. We use the voting ensemble rather than a single member because it offers a stronger and more stable representative of the classical model class. Details of the model configurations and hyperparameters are provided in Appendix Table~\ref{tab:hp}.

For deep learning, we employ SchNet~\cite{Schutt2017}, a continuous-filter convolutional neural network model that takes atomic numbers and Cartesian coordinates as input and learns atom-centred representations through six continuous-filter interaction blocks, followed by an atom-wise readout layer. Since both targets are intensive, so we use mean pooling for QM9 and Matbench MP. SchNet constructs the QM9 radius graph internally from the Cartesian coordinates with a cutoff of 6.0~\AA{}. For the materials in the Matbench MP dataset, neighbour lists with a cutoff of 5.0~\AA{} are constructed using ASE~\cite{Larsen2017} and supplied to a periodic variant of SchNet. Further details of the training procedure are provided in Appendix A. SchNet is not evaluated on the Pourbaix stability and QM9small datasets because of their relatively small size.

\subsection{Sampling strategies}

We considered a pool-based sampling framework in this work \cite{lewis1995sequential}. In this framework, an initial labelled set $\mathcal{L}^{(0)}$ of $N_0$ samples is drawn at random from the pool. At iteration $t$ the model is trained on $\mathcal{L}^{(t)}$, and a batch of $b$ samples is selected from the unlabeled pool $\mathcal{U}^{(t)}$, labeled, and added to $\mathcal{L}^{(t)}$~\cite{Settles2009}. All strategies and models share the same initial set within each repetition. Three acquisition strategies were compared:

\textbf{Random sampling (RS)} draws $b$ candidates uniformly and serves as the baseline.
\\
\textbf{Uncertainty sampling (US)} uses query-by-committee~\cite{Seung1992}, scoring each candidate by the variance of the committee predictions,
\begin{equation}
a_{\mathrm{US}}(x)=\frac{1}{M}\sum_{m=1}^{M}\bigl(\hat y_m(x)-\bar y(x)\bigr)^2,
\end{equation}
with $\bar y$ the committee mean. For the classical ML models, the committee consists of RF, KRR and XGB models with the disagreement among these different models providing an uncertainty signal rather than bootstrap resampling of a single model, which was used for SchNet ($M=3$ models were trained on bootstrap resamples of $\mathcal{L}^{(t)}$). For Matbench MP, the US committee used a reduced SchNet configuration for faster training (fewer filters, epochs and a smaller batch size than the main model).

\textbf{Diversity sampling (DS)} applies TypiClust~\cite{Hacohen2022} as implemented in scikit-activeml package ~\cite{Herde2025}. All samples are clustered with $k$-means~\cite{Lloyd1982} into $|\mathcal{L}^{(t)}|+b$ clusters; for each of the $b$ acquisitions, the largest cluster without labelled members is selected, and its most typical sample is queried, where typicality is the inverse mean Euclidean distance to the $k=5$ nearest neighbours~\cite{Cover1967} within the cluster,
\begin{equation}
\mathrm{Typ}(x)=\Bigl(\tfrac{1}{k}\sum_{x_i\in k\text{-NN}(x)}\lVert x-x_i\rVert_2\Bigr)^{-1}.
\end{equation}
Clustering operates in descriptor space for the classical ML models and in the learned, latent space for the SchNet model.

Initial set sizes, batch sizes and iteration counts for AL are listed in Table~\ref{tab:al}. For QM9 and Matbench MP, the budgets, i.e., final training-set sizes, were 52\,106 and 50\,000 labels, i.e. half of the training pool ($\approx$40\% of the full dataset); for the Pourbaix stability dataset, the entire pool (5,496 labels) was consumed. Ensemble experiments were repeated with five random seeds, each with its own initial set, shared across strategies within a run whereas SchNet experiments were run only once owing to computational cost. 


\begin{table}[b]
\caption{Data splits and AL settings: initial size $N_0$, batch size $b$, number of iterations $n_{\mathrm{AL}}$.}
\label{tab:al}
\begin{ruledtabular}
\begin{tabular}{lrrrrrr}
Dataset & Train & Val. & Test & $N_0$ & $b$ & $n_{\mathrm{AL}}$ \\
\hline
QM9          & 107,106 & 13,386 & 13,389 & 7,106 & 5\,000 & 9 \\
QM9small     & 5,356   & 670 & 669 & 356 & 500 & 10 \\
Matbench MP  & 106,201 & 13,276 & 13,275 & 5,000 & 5,000 & 9 \\
Pourbaix Stability & 5,496 & 687 & 687 & 496 & 500 & 10 \\
\end{tabular}
\end{ruledtabular}
\end{table}

The performance of sampling strategies is reported as test RMSE (primary metric), MAE and $R^2$. Following Li~\textit{et al.}~\cite{Li2023}, data efficiency is quantified by
\begin{equation}
\Nten=\min\{N:\ \mathrm{RMSE}(N)\le 1.1\times\RMSEfull\},
\end{equation}
where $\RMSEfull$ is the error of the model trained on the full pool. The crossing was obtained by linear interpolation on a logarithmic $N$ axis, counted only if the error stayed below the threshold thereafter, and is also given as a fraction of the pool. A smaller $\Nten$ than RS indicates high redundancy of the dataset for the given model and property. Because $\Nten$ is defined relative to each model's own baseline, cross-model comparisons use absolute errors.

\section{Results and Discussion}
\label{sec:results}

\subsection{Full-data baselines}

\begin{figure*}[t]
\centering
\begin{subfigure}{0.48\linewidth} 
    \centering
    \safefig{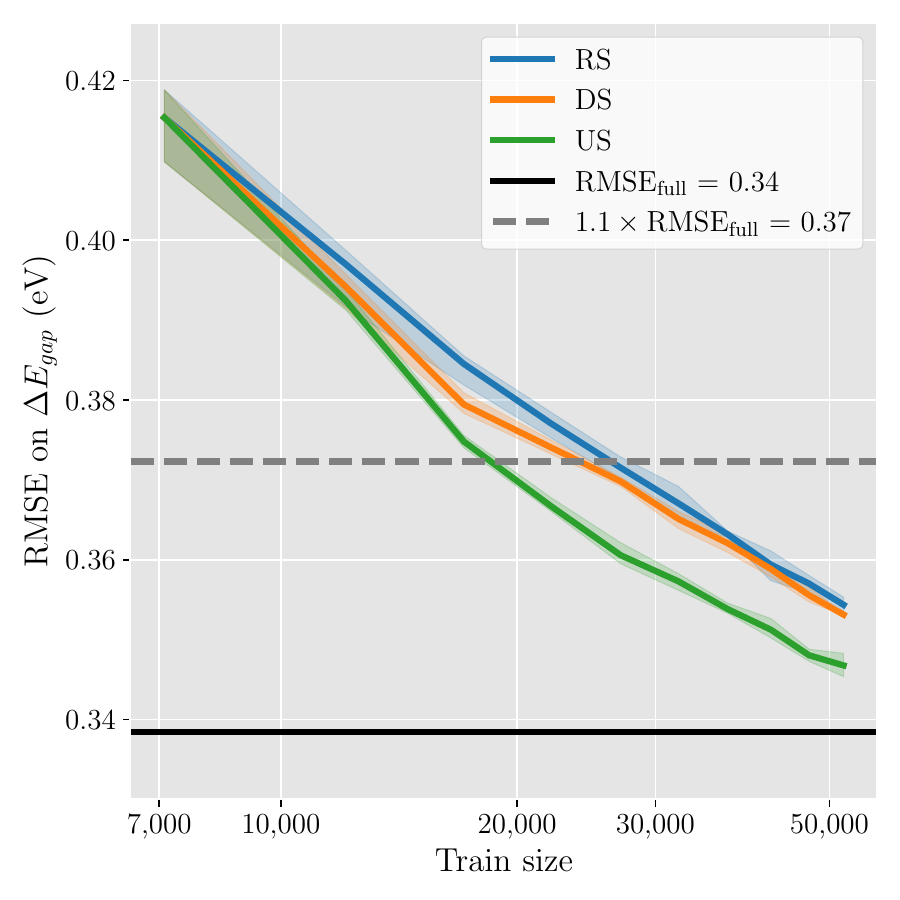}{\linewidth}
    \caption{QM9: Descriptor-based ensemble}
\end{subfigure}
\hfill 
\begin{subfigure}{0.48\linewidth} 
    \centering
    \safefig{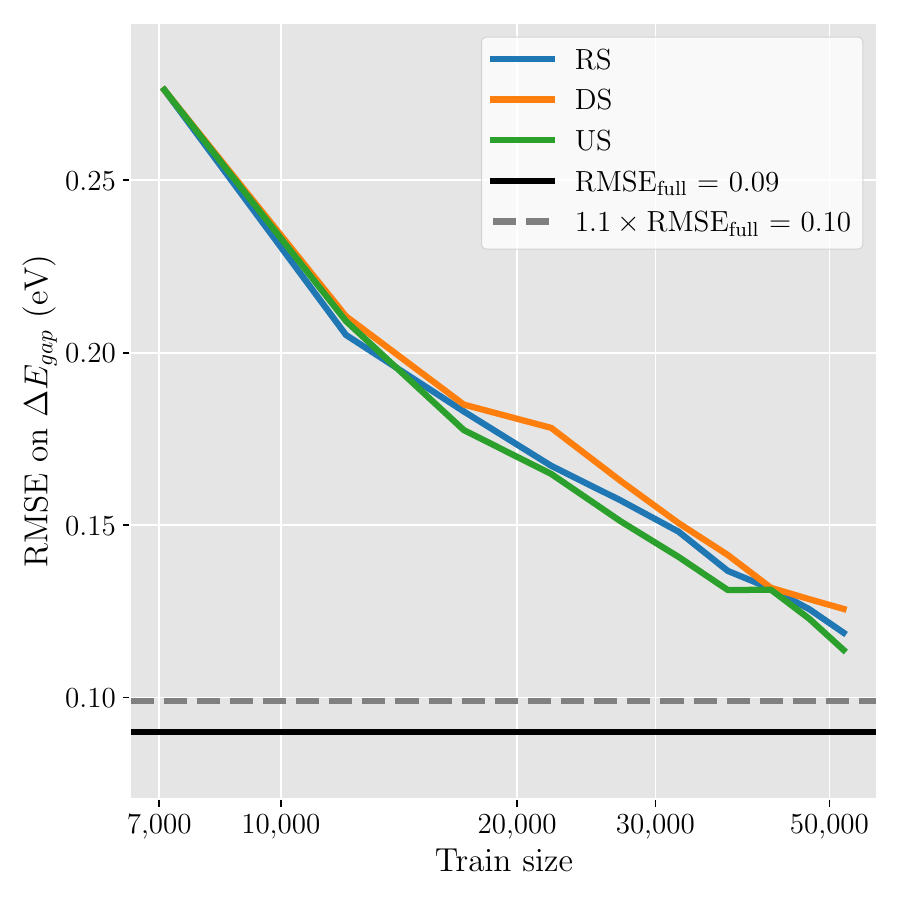}{\linewidth}
    \caption{QM9: SchNet}
\end{subfigure}
\caption{AL learning performance curves on QM9 for (a)~the descriptor-based ensemble of ML models and (b) SchNet. Curves show random sampling (RS), uncertainty sampling (US) and diversity sampling (DS). Solid black line indicates the RMSE obtained using the full dataset ($\RMSEfull$) and the dashed grey line indicates $1.1\times\RMSEfull$. The x-axis represents the number of samples in the training dataset and the y-axis represents the RMSE evaluated on a hold-out test set. The solid lines represent the mean values, and the shaded region in (a) spans the minimum--maximum range over five independent runs.}
\label{fig:qm9}
\end{figure*}

Table~\ref{tab:base} lists the test set performance for models trained on the full training data pool. SchNet outperforms the ensemble on both large datasets with a smaller RMSE of 0.090 vs.\ 0.339~eV for the HOMO-LUMO gap and 0.070 vs.\ 0.268~eV/atom for the formation energy prediction tasks. This can be attributed to the fact that the composition-derived descriptors carry no structural information and thus impose an information ceiling on a strongly structure-dependent property like formation energy, whereas SchNet encodes bond lengths, local coordination environments and atomic packing. The QM9small ensemble, trained on the 5,356-sample pool, reaches RMSE = 0.387~eV, slightly higher than the larger QM9 dataset. On the Pourbaix stability, the ensemble obtains a reasonable accuracy with RMSE = 0.243~eV, since $\Delta G_{pbx}$ is computed across different compositions present in the Pourbaix diagram of a material and hence compositional information alone aids the learning task.

\begin{table}[t]
\caption{Test-set performance of models trained on the full training pool.
RMSE and MAE are reported in the target property's units.}
\label{tab:base}
\begin{ruledtabular}
\begin{tabular}{llccc}
Dataset & Model & RMSE & MAE & $R^2$ \\
\hline
\multirow{2}{*}{QM9}
    & Ensemble & 0.339 & 0.257 & 0.931 \\
    & SchNet   & 0.090 & 0.062 & 0.995 \\
\hline
\multirow{2}{*}{QM9small}
    & Ensemble & 0.387 & 0.303 & 0.907 \\
    & SchNet   & \multicolumn{3}{c}{---} \\
\hline
\multirow{2}{*}{Matbench MP}
    & Ensemble & 0.268 & 0.140 & 0.947 \\
    & SchNet   & 0.070 & 0.036 & 0.996 \\
\hline
\multirow{2}{*}{Pourbaix Stability}
    & Ensemble & 0.243 & 0.158 & 0.993 \\
    & SchNet   & \multicolumn{3}{c}{---} \\
\end{tabular}
\end{ruledtabular}
\end{table}

\subsection{Active Learning Performance}

\begin{figure}[t]
\centering
\safefig{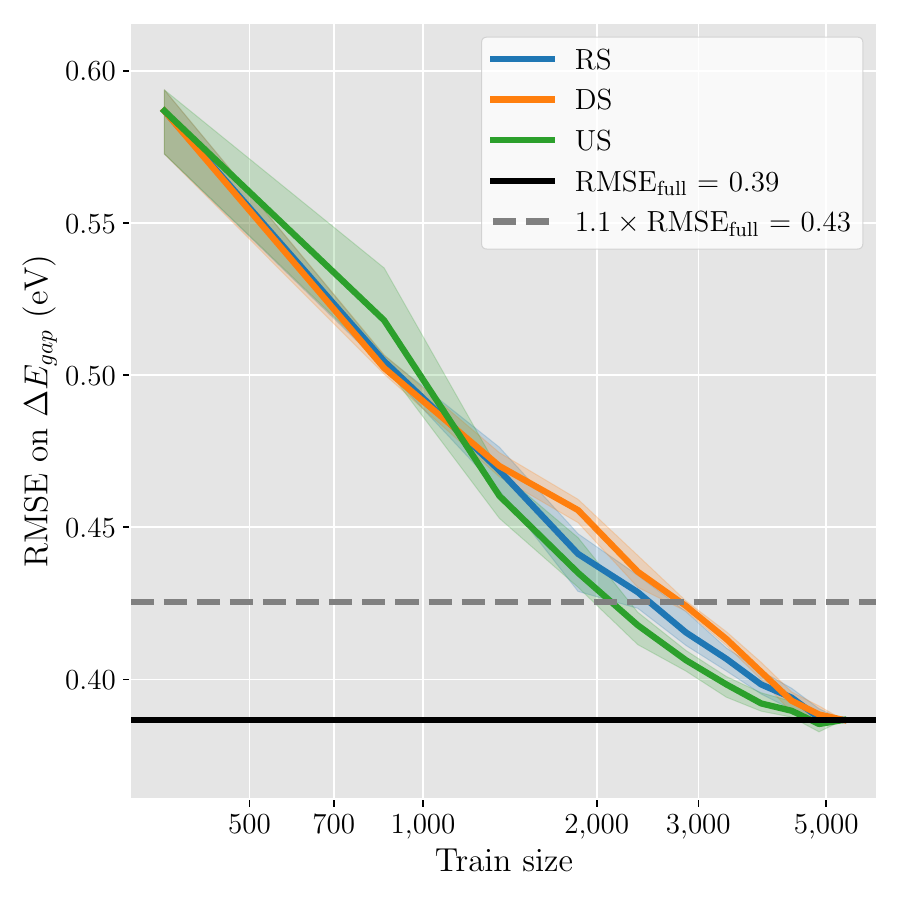}{\linewidth}
\caption{AL learning curves on QM9small for the descriptor-based ensemble; layout as in Fig.~\ref{fig:qm9}. The budget covers the full pool, so all strategies converge to the full-data limit.}
\label{fig:qm9small}
\end{figure}

\subsubsection{HOMO-LUMO Gap Prediction}
The AL performance of ensemble-based classical models and SchNet for HOMO-LUMO gap is shown in Fig.~\ref{fig:qm9}a \& b, respectively. In these plots, the RMSE evaluated on the hold-out set is compared with the number of samples acquired during AL iterations. For the ensemble-based models, all the sampling strategies start from a test RMSE of 0.415~eV at a training data size of 7\,106 and decrease monotonically, indicating improved performance with more training data. It can be seen that the performance curve of US lies below RS at every iteration, reaching the 10\% threshold at $\Nten\approx19\,000$ (17\% of the pool) compared with $\approx26\,000$ (25\%) for RS, and at the end of the budget (52\,106 labels), RMSE reaches 0.347~eV, compared with 0.354~eV for RS. DS occupies an intermediate position ($\Nten\approx24\,000$, 23\% of the pool; final RMSE = 0.353~eV), indicating that it is systematically better than RS in terms of $\Nten$, but the improvement is small and, at higher dataset sizes, is essentially indistinguishable from that of RS. The consistently observed higher performance of US indicates that disagreement among different regressors is a meaningful informativeness signal for the HOMO-LUMO gap, steering acquisition toward underrepresented regions of chemical space rather than merely toward already acquired labels. Coverage-oriented DS in descriptor space, which does not use any information about the model's current errors, captures only part of this benefit.

On the contrary to the ensemble, SchNet starts from a lower error (0.276 vs.\ 0.415~eV at 7\,106 labels) and improves by more than a factor of two within the budget (Fig.~\ref{fig:qm9}b). It is therefore more accurate than the ensembles even at the smallest budget. However, no sampling strategy reaches the 10\% threshold ($1.1\times\RMSEfull=0.099$~eV), and the lowest error achieved after full AL iterations (0.114~eV observed for US) remains 26\% above the full-data baseline. The performance curves are monotonic, with all three strategies decreasing at every cycle apart from a 0.0001~eV uptick for US at 42\,106 labels. US attains the lowest final error (0.114~eV), followed by RS (0.119~eV) and DS (0.126~eV); the three values differ by at most 0.012~eV, small relative to the $\approx$0.16~eV reduction achieved over the budget. Note that we fixed the number of AL iterations to nine for QM9 and Matbench MP, stopping at training-set sizes of 52\,106 and 50\,000 labels, respectively, rather than training until the complete dataset is scanned.

\begin{figure*}[htbp] 
\centering
\begin{subfigure}{0.48\linewidth}
    \centering
    \safefig{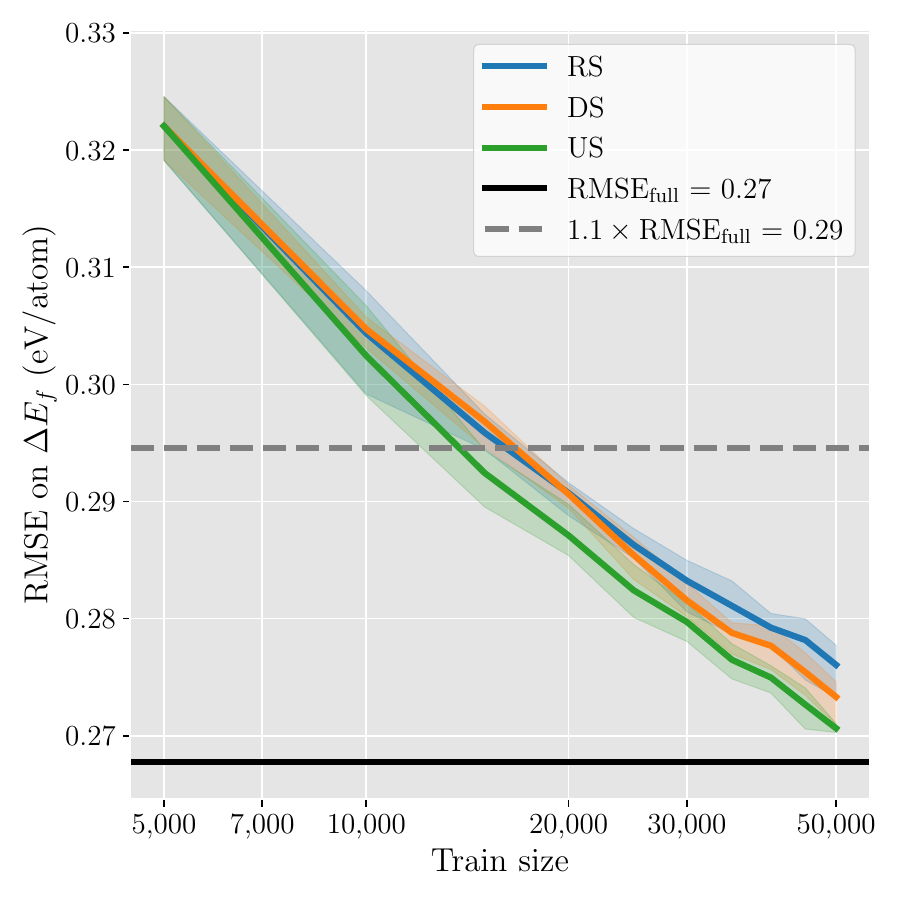}{\linewidth}
    \caption{Matbench MP: Descriptor-based ensemble}
\end{subfigure}
\hfill 
\begin{subfigure}{0.48\linewidth}
    \centering
    \safefig{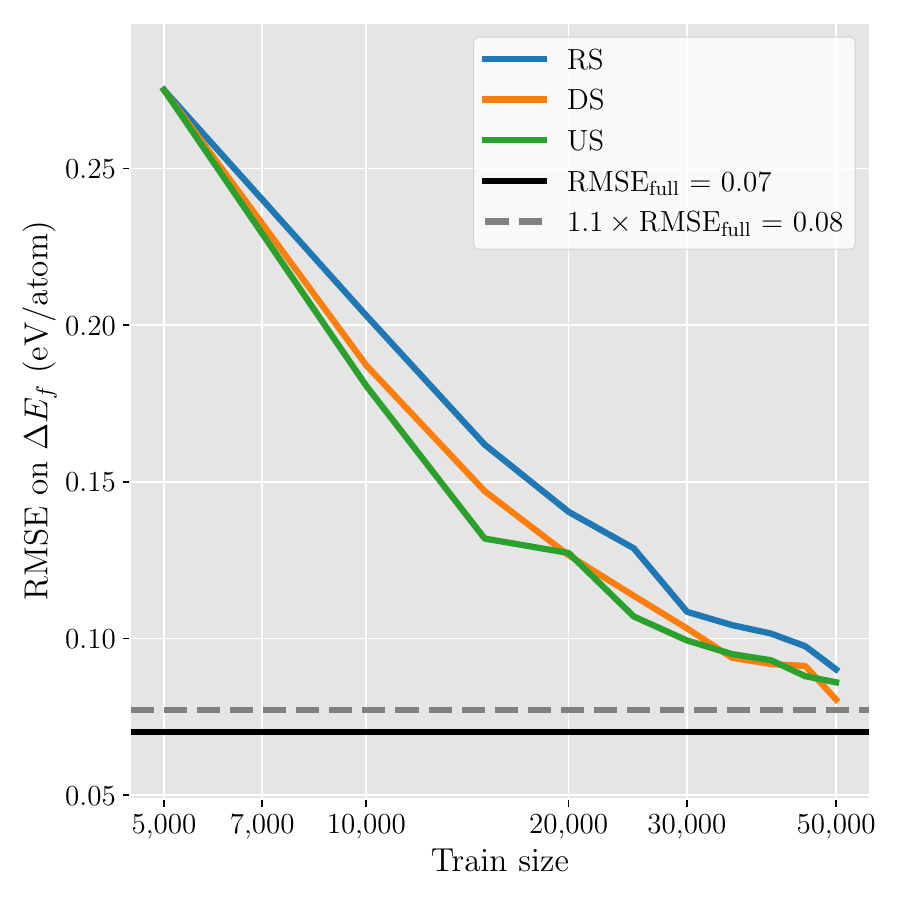}{\linewidth}
    \caption{Matbench MP: SchNet}
\end{subfigure}
\caption{AL learning performance curves for formation energy prediction task based on the Matbench MP dataset, layout as in Fig.~\ref{fig:qm9}.}
\label{fig:mp}
\end{figure*}

\textbf{Effect of pool size.} To analyse the effect of pool size for a fixed property and model, we repeated the AL experiment on QM9small, a random 5\% subset of QM9 (6\,695 molecules; 5\,356 in the training pool), using the descriptor-based ensemble only. The full-data baseline is 0.387~eV (Table~\ref{tab:base}) and the AL settings are listed in Table~\ref{tab:al}. All strategies start from 0.587~eV and converge to 0.387~eV at 5\,356 labels (Fig.~\ref{fig:qm9small}). Because the budget covers the full pool, all strategies end at the same accuracy; their benefit therefore lies in reaching the threshold with fewer labels, not in a better final model. US reaches the 10\% threshold first, at $\Nten\approx 2\,100$ labels (40\% of the pool), compared with $\approx 2\,500$ (46\%) for RS and $\approx2\,800$ (52\%) for DS. US therefore saves $\approx 400$ labels relative to RS ($\approx$16\% of the RS budget; the threshold fraction drops from 46\% to 40\%), while DS requires $\approx 300$ labels more than RS. US is the most data-efficient strategy, followed by RS and DS. This differs from the full QM9 dataset, where DS was slightly better than RS, consistent with the smaller pool offering less redundancy for descriptor-space DS to exploit. It should be also noticed that the threshold fractions are larger than those on QM9 dataset (40--52\% vs. 17--25\%). 

\subsubsection{Formation Energy Prediction}

In the case of the formation energy prediction task based on the Matbench MP dataset, US reaches the 10\% threshold at $\Nten\approx14\,000$ labels (13\% of the pool), RS at 16\,000 (15\%) and DS at 17\,000 (16\%) (Fig.~\ref{fig:mp}a). This clearly shows the very high redundancy of this dataset, which is expected because of very similar inorganic crystal structures present in MaterialsProject datasets from high-throughput simulations. At the end of the budget, all curves are within $\sim 3\%$ of the full-data baseline, and the mean RMSE increases in the order US, DS, RS (0.271, 0.273 and 0.276~eV/atom, respectively). The enhanced early performance of US suggests that querying models for highly uncertain configurations effectively isolates and corrects the most critical knowledge gaps in the materials space early in the training process. Interestingly, the lower final error of DS compared with RS emerges only at larger training-set sizes, although DS reaches the 10\% threshold slightly later (17\,000 vs.\ 16\,000 labels; Table~\ref{tab:eff}): DS is marginally worse than RS up to 15\,000 labels, essentially indistinguishable at 20\,000, and lower on average from 25\,000 labels onward, with the gap growing to $\approx$0.003~eV/atom at 50\,000. This gradual separation indicates that at lower data regimes, RS provides sufficiently broad phase-space coverage. However, as the dataset grows, DS begins to outpace RS by systematically avoiding the redundant sampling of overrepresented crystal structures, ensuring a more efficient exploration of the remaining chemical space.  

The performance of the SchNet model on Matbench MP shows a different pattern compared to the QM9 dataset (Fig.~\ref{fig:mp}b). Starting from 0.275~eV/atom at 5\,000 labels, US is observed to show the highest performance through most of the AL iterations, except at the later stage, where it converges to similar performance to DS. It is to be noted that DS outperforms RS at every subsequent iteration, ending at 0.080 versus 0.090~eV/atom (11\% lower error). No strategy reaches the threshold of 0.077~eV/atom within the 50\,000-label budget, although DS comes closest. These results indicate that the relative efficacy of AL strategies is strongly data-dependent, meaning a strategy optimized for one dataset may not generalise as the most effective choice for another.



\begin{table*}[t]
\caption{Data efficiency $\Nten$ for random sampling (RS), uncertainty sampling (US) and diversity sampling (DS) sampling strategies for all the datasets considered in this study. The fraction of the training pool is indicated in parentheses}
\label{tab:eff}
\begin{ruledtabular}
\begin{tabular}{llccc}
Dataset & Model & RS & US & DS \\
\hline
QM9              & Ensemble & 26\,000 (25\%) & \textbf{19\,000 (17\%)} & 24\,000 (23\%) \\
QM9small         & Ensemble & 2\,500 (46\%)   & \textbf{2\,100 (40\%)}   & 2\,800 (52\%) \\
Matbench MP      & Ensemble & 16\,000 (15\%) & \textbf{14\,000 (13\%)} & 17\,000 (16\%) \\
Pourbaix stability & Ensemble & 4\,200 (77\%)  & \textbf{3\,000 (54\%)}  & 3\,600 (66\%) \\
\end{tabular}
\end{ruledtabular}
\vspace{0.5ex}
\begin{minipage}{\linewidth}
\footnotesize
\textit{Note:} $\Nten$ is obtained by linear interpolation of the mean learning curve and rounded to the nearest 1\,000 labels (100 for QM9small and Pourbaix stability). Percentages are relative to the training pools in Table~\ref{tab:al} and are computed from the unrounded crossings, so subtracting the rounded $\Nten$ values may differ slightly.
\end{minipage}
\end{table*}

\subsubsection{Pourbaix Decomposition Energy Prediction}

\begin{figure}[t]
\centering
\safefig{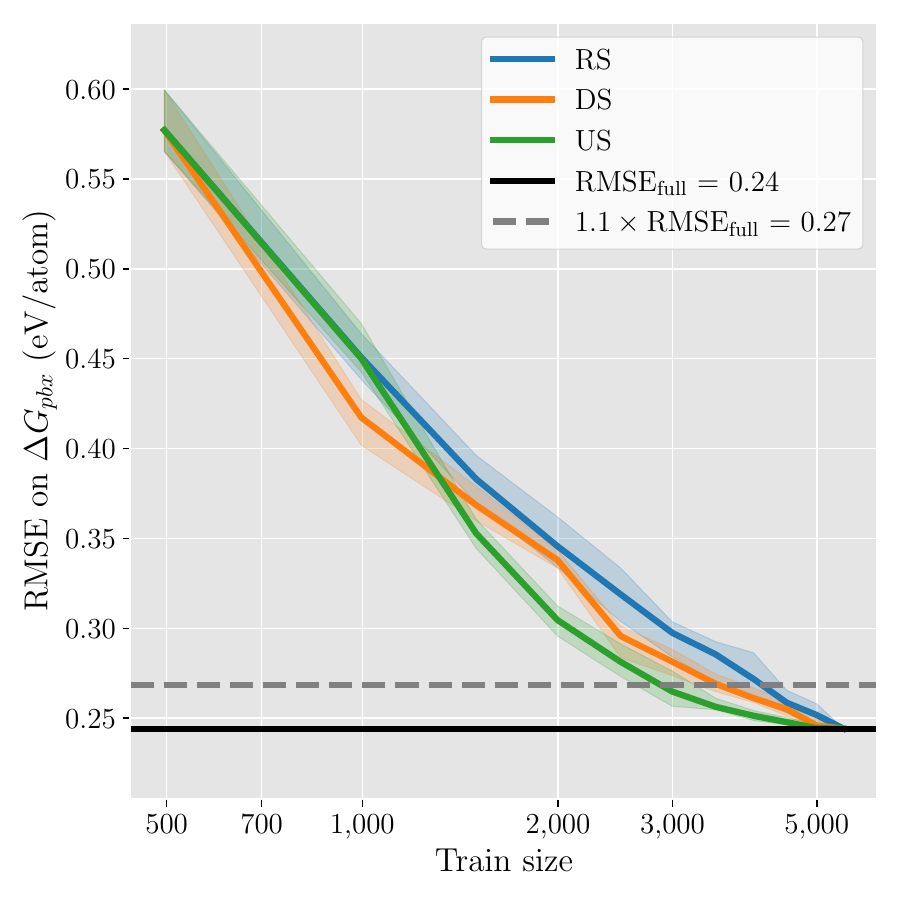}{\linewidth}
\caption{AL learning performance curves on Poubrbaix stability dataset for the ensemble; layout as in Fig.~\ref{fig:qm9}. The budget covers the full pool, so all strategies converge to the full-data limit.}
\label{fig:hse}
\end{figure}

Since the pool is small (5,496 labels) and structural relaxations were not performed at the HSE06 level for curating this dataset, we did not evaluate SchNet on this dataset, and hence the comparison is limited to sampling strategies for the ensemble. Here, the differences between strategies are the largest among the three properties considered when expressed as a fraction of the pool (Fig.~\ref{fig:hse}). US reaches the 10\% threshold at $\approx$3\,000 samples (54\% of the pool), DS at $\approx$3\,600 (66\%) and RS at $\approx$4\,200 (77\%). US thus saves $\approx$1\,200 labels, or about 22\% of the pool, relative to RS. We extended the sampling to the full pool, where all strategies converge to the full-data limit (5\,496 labels). As observed for the QM9small dataset, this benefit lies in reaching the threshold earlier rather than in a better final model (Table~\ref{tab:eff}). This is particularly important here because for a property with expensive labels (HSE06-level calculations), reaching a target accuracy earlier translates directly into reduced computational cost.


\subsection{Dataset Redundancy Analysis}

To further assess the performance of various strategies across all the datasets, we list their corresponding $\Nten$ values in Table~\ref{tab:eff} for the ensemble. It is particularly important to note that, even for RS, only 15--25\% of the pool is sufficient on QM9 and Matbench MP datasets, so that $\gtrsim$75\% of the entries contribute little additional accuracy. This agrees with the finding of Li~\textit{et al.} that only about 13--30\% of Materials Project entries are informative for composition-based tree models at the same threshold~\cite{Li2023}. QM9small, a 5\% subset of QM9, requires $\approx$46\% of its pool under RS, between the two large datasets and the Pourbaix stability dataset. For the $\Delta G_{pbx}$ prediction,  RS needs $\approx$77\% of the pool (4\,200 samples). This sharp contrast can be attributed to the significantly smaller overall size and lower inherent redundancy of the HSE06 dataset compared to the massive scale of QM9 and Matbench MP. Because the total available pool for the Pourbaix stability dataset is highly constrained, RS struggles to capture the full structural diversity without exhausting the majority of the available data. However, it is precisely in this data-scarce regime that AL demonstrates its greatest value. Relative to RS, US reduces the threshold fraction from 77\% to 54\%, reaching the target with 3\,000 labels instead of 4\,200 (Table~\ref{tab:eff}). This emphasizes that when data are limited and computationally expensive to acquire, as is typical for high-fidelity properties like DFT-HSE06, uncertainty-driven sampling is crucial for maximizing model accuracy with a minimal number of calculations.



\subsection{Effect of Choice of Surrogate Models}

\begin{figure*}[htbp]
\centering
\begin{subfigure}{0.48\linewidth}
    \centering
    \safefig{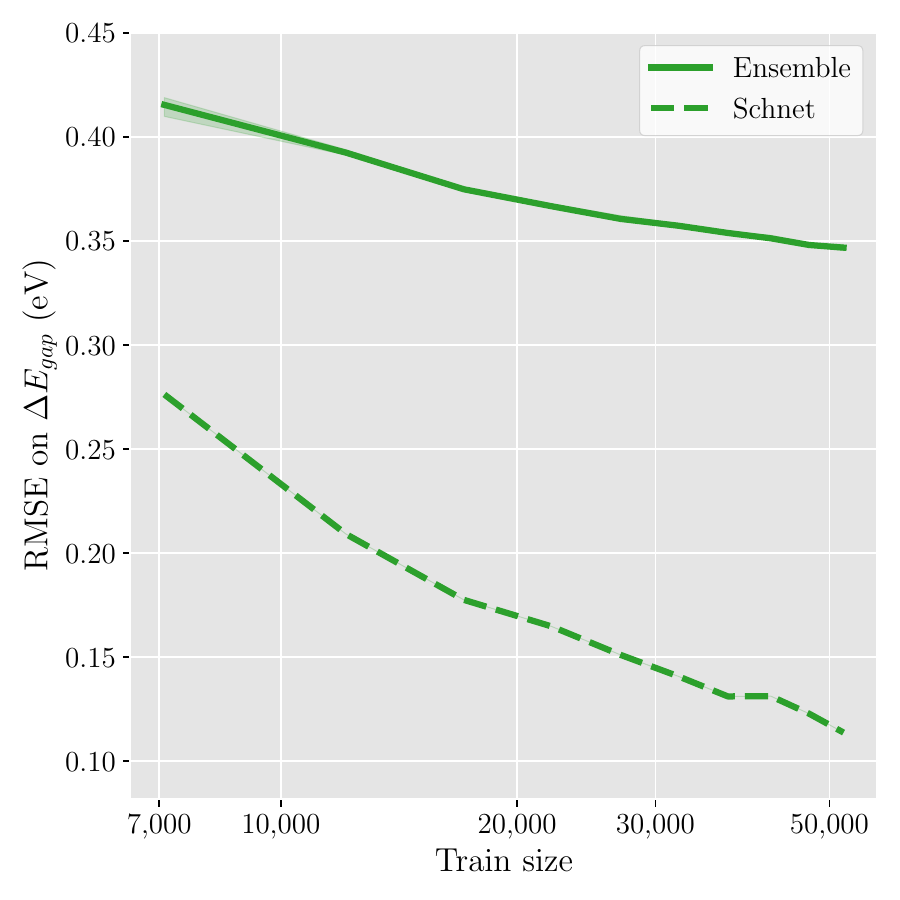}{\linewidth}
    \caption{QM9: Ensemble vs. SchNet}
\end{subfigure}
\hfill
\begin{subfigure}{0.48\linewidth}
    \centering
    \safefig{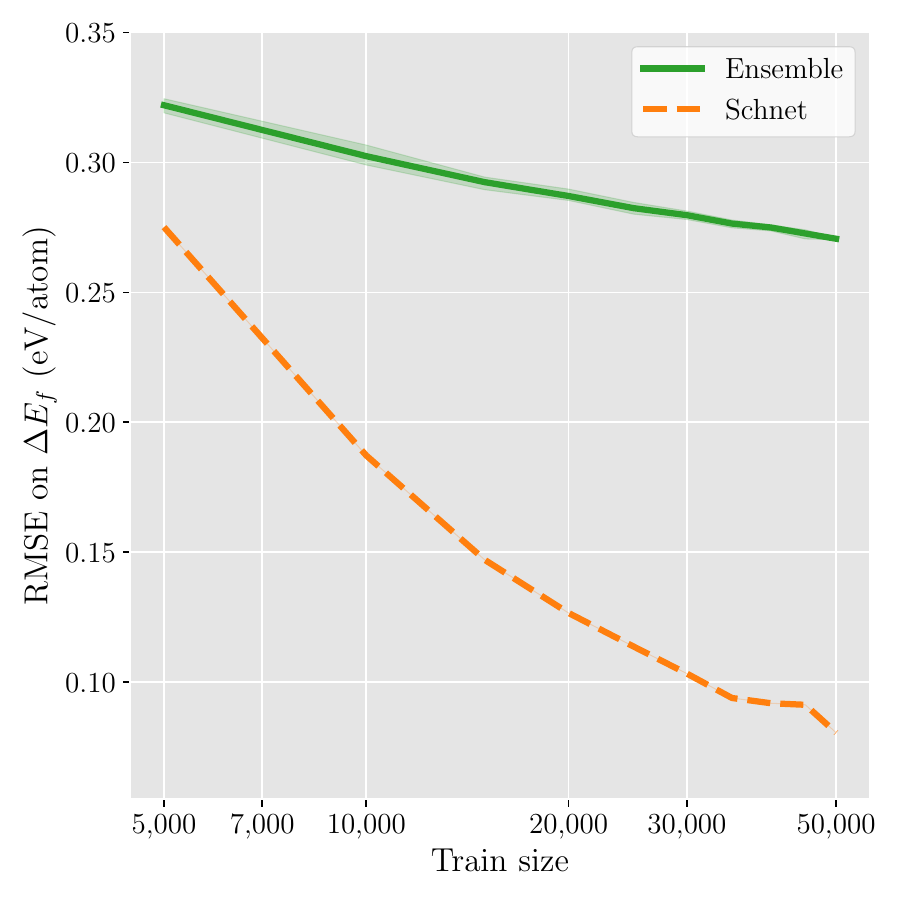}{\linewidth}
    \caption{Matbench MP: Ensemble vs. SchNet}
\end{subfigure}
\caption{Ensemble vs.\ SchNet with the best strategy per dataset and model: (a)~QM9 (uncertainty sampling for both); (b)~Matbench MP (uncertainty sampling for the ensemble and diversity sampling for SchNet).}
\label{fig:comp}
\end{figure*}

Herein, we investigate the effect of model choice (ensemble vs SchNet) on the performance of AL performance for the same dataset. 
Because $\Nten$ is defined relative to each model's own full-data error, the threshold is lower in absolute terms for the more accurate model. The SchNet curves never reach their threshold within the budget and hence $\Nten$ could not be determined for SchNet. It is important to note that $\Nten$ values are not directly comparable across model classes. We therefore compare the models using absolute errors (Fig.~\ref{fig:comp}).
For the QM9 dataset, SchNet is superior at every budget tested: it is already better at the initial 7\,106 labels (0.276 vs.\ 0.415~eV) and the gap widens to 0.114 vs.\ 0.347~eV at 52\,106 labels. On Matbench MP, SchNet is also superior at all budgets tested, and the gap widens to roughly a factor of three by the end of the 50\,000-label budget. The ensemble is ultimately limited by its fixed descriptors with the composition-derived features that have limited representation capability for the structural information that SchNet learns, and for QM9 the fingerprints and scalar descriptors do not capture the HOMO-LUMO gap as accurately as the learned representation. These results offer practical guidelines for AL deployments. When full structural information is unavailable, the descriptor-based ensemble combined with US is the most robust option tested. When structural data are available, SchNet is the more accurate model class at all budgets tested on both QM9 and Matbench MP; the ensemble is computationally cheaper but does not provide a low-budget accuracy advantage. 


\subsection{Limitations}
\label{sec:limitations}
Due to GPU availability constraints, the SchNet models were trained using a single run per configuration. Consequently, we are unable to report standard deviations for the SchNet learning curves, which hinders a rigorous assessment of the model's statistical variance across different random weight initializations.


\section{Conclusions}
\label{sec:conclusions}

We evaluated active learning (AL) strategies for curating training data across molecular and materials datasets, using both a descriptor-based ensemble and the structure-aware SchNet model. Our results show that how much training data is needed, and how much a well-chosen selection strategy can save, are not fixed properties of a dataset but depend on the interplay of dataset size, target property, model and representation. For the descriptor-based models, only 15--25\% of the two large datasets (QM9 and Matbench MP) was needed to come within 10\% of the full-data error under random sampling, whereas the smaller datasets required 46--77\%. Within this picture, uncertainty sampling was the most label-efficient strategy for the descriptor-based models on all four datasets, reducing the number of labels required by roughly 13--30\% relative to random sampling, with the largest gain on the smallest and most expensive dataset. Diversity sampling, by contrast, gave no consistent advantage, and the best strategy for SchNet changed with the dataset. Moreover, the accuracy of the model at the start of AL strongly shaped its performance throughout the sampling, and the model and its input representation often mattered more than the choice of sampling strategy. Building an efficient training set therefore requires balancing these factors rather than applying a single recipe. Future work will focus on a benchmarking platform that offers the community an expanded suite of ML models, sampling strategies and performance metrics for quantifying the redundancy of target properties.

\begin{acknowledgments}
A.S.N thanks the German Research Foundation (DFG) for their support through the Walter-Benjamin Fellowship Program (project No. 540316537). 

\end{acknowledgments}

\section*{Data and code availability}
The code used in this work is available at \url{https://github.com/minhnghian03/property-prediction-al}. The datasets are publicly available from the cited sources.

\appendix

\section{Data splits and training protocols}
\label{app:training}

\subsection{Data splits}

Each dataset was divided into 80\% training, 10\% validation and 10\% test subsets, stratified by quantile bins of the target property; the resulting sizes are listed in Table~\ref{tab:al}. The training subset defines the AL pool, the validation subset is used for early stopping, and the test subset is held out for all reported metrics. Hyperparameters were selected by five-fold cross-validation on the training subset.

\subsection{Classical ensemble}
\label{app:ensemble}

The classical model is an unweighted voting ensemble of RF, KRR and XGB. Continuous descriptors were standardized with a scaler fitted on the training subset, while fingerprint features were used without scaling. KRR used a Laplacian kernel, with the Nystr\"om approximation (2\,500 components) for QM9 and Matbench MP, and exact KRR for HSE06 and QM9small. The ensemble prediction is the mean of the three members. Table~\ref{tab:hp} lists the selected hyperparameters.

Targets were used in their native units: the QM9 and QM9small gap in Hartree and the formation and Pourbaix decomposition energies in eV/atom. QM9 predictions were converted to eV for all reported metrics.

\begin{table*}[t]
\caption{Hyperparameters of the ensemble members (QM9, QM9small, Matbench MP and
Pourbaix stability), selected by five-fold cross-validation on the training
subset.}
\label{tab:hp}
\small
\begin{ruledtabular}
\begin{tabular}{llcccc}
Model & Parameter & QM9 & QM9small & MP & HSE06 \\
\hline
RF  & n\_estimators & 100 & 100 & 500 & 500 \\
    & max\_features & sqrt & sqrt & 0.8 & 0.8 \\
    & min\_samples\_leaf & 2 & 2 & 1 & 1 \\
KRR & kernel & Laplacian & Laplacian & Laplacian & Laplacian \\
    & $\alpha$ & 0.01 & 0.01 & 0.01 & 0.01 \\
    & $\gamma$ & 0.01 & 0.01 & 0.01 & 0.01 \\
    & Nystr\"om components & 2500 & --- & 2500 & --- \\
XGB & n\_estimators & 100 & 100 & 500 & 500 \\
    & learning\_rate & 0.1 & 0.1 & 0.1 & 0.1 \\
    & max\_depth & 4 & 4 & 4 & 4 \\
    & colsample\_bytree & 0.6 & 0.6 & 0.8 & 0.8 \\
    & min\_child\_weight & 3 & 3 & 3 & 3 \\
    & reg\_alpha & 0.1 & 0.1 & 1.0 & 1.0 \\
    & reg\_lambda & 1.0 & 1.0 & 1.0 & 1.0 \\
    & subsample & 0.8 & 0.8 & 0.6 & 0.6 \\
\end{tabular}
\end{ruledtabular}
\end{table*}

\subsection{SchNet}
\label{app:schnet}

SchNet was trained with the MSE loss, an initial learning rate of $10^{-4}$, and mini-batches of 32 graphs for at most 150 epochs; the optimizer, weight decay and early-stopping settings are listed in Table~\ref{tab:schnet}. The learning rate was halved after 5 epochs without validation improvement (minimum $10^{-6}$), and early stopping used a patience of 20 epochs with the best-validation weights restored. QM9 used mean readout and an internal radius graph with a 6.0~\AA{} cutoff. Matbench MP used mean readout and periodic neighbour lists with a 5.0~\AA{} cutoff constructed with ASE. QM9 targets were used in Hartree and converted to eV for reporting; Matbench MP targets were standardized to zero mean and unit variance before training and converted back to eV/atom for evaluation. Table~\ref{tab:schnet} lists the architecture and training hyperparameters.

\begin{table}[t]
\caption{SchNet architecture and training hyperparameters.}
\label{tab:schnet}
\begin{ruledtabular}
\begin{tabular}{lcc}
Parameter & QM9 & Matbench MP \\
\hline
hidden channels & 64 & 64 \\
filters & 256 & 128 \\
interaction blocks & 6 & 6 \\
Gaussians & 50 & 50 \\
cutoff (\AA) & 6.0 & 5.0 \\
readout & mean & mean \\
optimizer & AdamW & AdamW \\
learning rate & $10^{-4}$ & $10^{-4}$ \\
weight decay & $10^{-6}$ & $10^{-6}$ \\
batch size & 32 & 32 \\
max.\ epochs & 150 & 150 \\
early-stopping patience & 20 & 20 \\
LR-scheduler patience & 5 & 5 \\
\end{tabular}
\end{ruledtabular}
\end{table}

\subsection{Implementation and hardware}

The implementation used Python 3.13.9, scikit-learn 1.7.2, XGBoost 3.2.0, PyTorch 2.12.0 (CUDA 12.6), PyTorch Geometric 2.8.0 with the torch-scatter, torch-sparse extensions and pyg-lib, Skorch 1.3.1, RDKit 2026.3.1, matminer 0.10.0, ASE 3.26.0 and scikit-activeml 1.0.0. SchNet was trained on NVIDIA GeForce RTX 3050, GeForce GTX 1080 and Quadro RTX 5000 Max-Q GPUs, and the ensembles on Intel Xeon Gold 6526Y CPUs.


\end{document}